\documentclass[prl,twocolumn,nofootinbib,superscriptaddress]{revtex4-1}
\usepackage{amssymb}
\usepackage{amsmath}
\usepackage{amstext}
\usepackage{amsfonts}
\usepackage{bbold}
\usepackage{slashed}
\usepackage{color}
\usepackage{microtype}
\usepackage{graphicx}

\def\dd{{\rm d}}
\def\vphi{\varphi}

\def\be{\begin{equation}}
\def\ee{\end{equation}}

\newcommand{\rs}{r_s}
\newcommand{\rp}{r_+}
\newcommand{\SL}{${\rm sl}(2,\mathbb{R})$}

\definecolor{darkblue}{rgb}{0.15,0.35,0.55}
\definecolor{reddish}{rgb}{0.65, 0.2, 0.2}

\usepackage[linktocpage=true]{hyperref}
\hypersetup{
colorlinks=true,
citecolor=darkblue,
linkcolor=reddish,
urlcolor=darkblue,
pdfauthor={},
pdftitle={},
pdfsubject={}
}

\definecolor{darkgreen}{RGB}{50,150,0}

\allowdisplaybreaks

\begin{document}

\title{Near-Zone Symmetries of Kerr Black Holes}

\author{Lam Hui}
\affiliation{Center for Theoretical Physics, Department of Physics,\\
 Columbia University, New York, NY 10027, USA}
\author{Austin Joyce}
\affiliation{Department of Astronomy and Astrophysics,
University of Chicago, Chicago, IL 60637, USA}
\author{Riccardo Penco}
\affiliation{Department of Physics, Carnegie Mellon University, Pittsburgh, PA 15213, USA}
\author{Luca Santoni}
\affiliation{ICTP, International Centre for Theoretical Physics, Strada Costiera 11, 34151, Trieste, Italy}
\author{Adam R. Solomon}
\affiliation{Department of Physics and Astronomy, McMaster University, Hamilton, ON, L8S 4M1, Canada}
\affiliation{Perimeter Institute for Theoretical Physics, Waterloo, ON, N2L 2Y5, Canada}

\begin{abstract}
\noindent
We study the near-zone symmetries of a massless
scalar field on four-dimensional black hole backgrounds. We provide a geometric
understanding that unifies various recently discovered symmetries as part of an ${\rm SO}(4,2)$ group. Of
these, a subset are exact symmetries of the static sector and give
rise to the ladder symmetries responsible for the vanishing of Love
numbers.  In the Kerr case, we compare different near-zone approximations in the literature, and focus on the implementation that retains the symmetries of the static limit.
We also describe the relation to spin-1 and 2 perturbations.
\end{abstract}  

\maketitle
 
\noindent
{\bf Introduction:}  Black hole perturbation theory has a long
history dating back to the work of Regge and Wheeler~\cite{Regge:1957td} and
Zerilli~\cite{Zerilli:1970se}.
Interestingly, recent investigations suggest the subject has
depths yet to be plumbed. A case in point is a number of symmetries
discovered in the past year~\cite{Charalambous:2021kcz,Hui:2021vcv}, which shed light on the well-known
vanishing of black hole Love numbers (characterizing a
black hole's static, non-dissipative tidal response)~\cite{Fang:2005qq,Damour:2009vw,Binnington:2009bb, Damour:2009va, Kol:2011vg, LeTiec:2020spy,LeTiec:2020bos,Chia:2020yla,Hui:2020xxx,Charalambous:2021mea}. 
In this paper, we
present a synthesis of these symmetries, and show how they fit
within a larger group containing further symmetries. 
Some of these
are familiar symmetries of the exact dynamics. 
The rest are
approximate symmetries in the low frequency regime. Of these, a subset
are exact symmetries of the static sector, and give rise to the ladder symmetries
discussed in \cite{Hui:2021vcv}. 
To keep the discussion simple, we focus largely on symmetries of a
massless scalar, first on a Schwarzschild, then Kerr, background. The connection
to spin-1 and 2 perturbations, via a spin ladder, is discussed in the Supplemental Material.

\vskip5pt
\noindent
{\bf  Effective near-zone metric:} 
We begin by considering the Schwarzschild case. Our starting point is
a free massless scalar field $\phi$
on a fixed 4D Schwarzschild background:
\be
	\dd s^2 = - f(r) \dd t^2 + \frac{\dd r^2}{f(r)} + r^2 \dd \Omega^2_{S^2} ,~\, f(r) = 1 - \frac{\rs}{r},
\ee
with $\dd\Omega_{S^2}^2\equiv \dd\theta^2+\sin^2\theta \, \dd\varphi^2$ and where $r_s\equiv2GM$ is the Schwarzschild radius.
The scalar's action can be written explicitly~as
\be
 \label{scalar action}
	S = \frac{1}{2} \int \dd t \dd r \dd \Omega\left[  \frac{r^4}{\Delta} (\partial_t \phi)^2 - \Delta (\partial_r \phi)^2  + \phi \nabla^2_\Omega  \phi \right] , 
\ee
with $\Delta (r) \equiv r ( r - r_s)$ and $\nabla^2_\Omega \equiv (1/ \sin\theta)\partial_\theta
( \sin\theta \,  \partial_\theta) + (1 / \sin^2\theta) \partial^2_\varphi$. 
In frequency space ($\phi \propto e^{-i\omega t}$), we wish to focus on
long-wavelength perturbations satisfying
$r_s \ll 1/\omega$.
The
behavior of $\phi$ in the \emph{near-zone} region defined by $r_s \leq r \ll
1/\omega$ is described by approximating the coefficient in front
of the kinetic term as follows: $r^4 / \Delta(r) \simeq r_s^4 /
\Delta(r)$~\cite{1973JETP...37...28S,1974JETP...38....1S}. 
Doing so means in the $\phi$ equation of motion, the time derivative
term $(r^4 / \Delta)\partial_t^2\phi$ is replaced by $(r_s^4 /
\Delta)\partial_t^2\phi$. This has the virtue of preserving the correct
singularity as $r \rightarrow r_s$, while still accurately capturing the dynamics at larger $r$, as long as $\omega r \ll1$.\footnote{These criteria do not uniquely fix the near-zone approximation for finite $\omega$. Nevertheless, the near-zone approximation of Schwarzschild we use is standard in the literature. The existence of the symmetries that we will discuss below can be viewed as an \emph{a posteriori} motivation for this particular implementation.}

\vskip3pt
In this limit, the action \eqref{scalar action} is the same as that of a massless scalar minimally coupled to an \emph{effective near-zone metric}: 
\be
\label{near zone metric}
	\dd s^2_{\text{near-zone}} = - \frac{\Delta}{r_s^2} \dd t^2 + \frac{r_s^2 }{\Delta} \dd r^2 + r_s^2 \dd \Omega^2_{S^2}.
\ee
In the static limit ($\omega=0$) the scalar behaves identically on
this metric as on the original Schwarzschild background. Nevertheless, it is advantageous to
work with the near-zone geometry, both because it allows us to go beyond the strictly static sector, and
because it has a richer symmetry structure. In fact, the metric
\eqref{near zone metric} is that of ${\rm AdS}_2\times S^2$. (The $(t,r)$ coordinates are a somewhat nonstandard covering of a portion of ${\rm AdS}_2$, which we describe in the Supplemental Material.) This immediately implies that the near-zone metric~\eqref{near zone metric} has 6 Killing vectors (KVs), in contrast to Schwarzschild, which has only 4.

\vskip3pt
Another advantage of the near-zone metric is that it describes a conformally
flat spacetime, unlike Schwarzschild. This implies that the metric~\eqref{near zone metric} has 9 additional conformal Killing vectors (CKVs). The near-zone metric
also has a vanishing Ricci scalar (though not Ricci tensor) because the curvature radii of ${\rm AdS}_2$ and $S^2$ are identical. This means that the scalar $\phi$ is effectively conformally coupled, guaranteeing the CKVs generate symmetries of the action in the near zone. 
We now turn to the study of these symmetries and their
physical consequences. 

\vskip5pt
\noindent
{\bf Near-zone symmetries:} The Killing vectors of ${\rm AdS}_2\times S^2$ in $(t,r,\theta, \varphi)$ coordinates are
\begin{subequations} \label{KVs}
\begin{align}
	T &= 2 r_s \, \partial_t , \\
	L_{\pm} &= e^{\pm t/2r_s} ( 2r_s \, \partial_r \sqrt{\Delta}\partial_t  \mp \sqrt{\Delta} \partial_r ) , \label{KVs-L}\\
	J_{23} &= \partial_\varphi , \\
	J_{12} &= \cos \varphi \, \partial_\theta - \cot \theta \sin \varphi \, \partial_\varphi , \\
	J_{13} &= \sin \varphi \, \partial_\theta + \cot \theta \cos \varphi \, \partial_\varphi .
\end{align}
\end{subequations}
The Killing vectors $L_0\equiv T$ and $L_\pm$ were first introduced in~\cite{Bertini:2011ga} and coincide with the zero-spin limit of the symmetries  discovered for Kerr in~\cite{Charalambous:2021kcz}. More recently they were encountered in the context of rotating STU supergravity black holes~\cite{Cvetic:2021vxa}.

\vskip3pt
The near-zone metric~\eqref{near zone metric} also possesses 9 conformal Killing vectors: 
\begin{subequations} \label{CKVs}
\begin{align}
	J_{01} &= - \tfrac{2 \Delta}{r_s} \cos \theta \, \partial_r - \tfrac{\partial_r \Delta}{r_s} \sin \theta \,\partial_\theta , \\
	J_{02} &= - \cos \varphi  \left[ \tfrac{2 \Delta}{r_s} \sin \theta \, \partial_r + \tfrac{\partial_r \Delta}{r_s} \left( \tfrac{\tan\varphi}{\sin\theta} \partial_\varphi - \cos \theta \partial_\theta \right) \right], \\
	J_{03} &= - \sin \varphi \left[ \tfrac{2 \Delta}{r_s}  \sin \theta \, \partial_r - \tfrac{\partial_r \Delta}{r_s} \left( \tfrac{\cot\varphi}{\sin\theta} \partial_\varphi + \cos \theta \partial_\theta \right) \right], \\
	K_{\pm} &= e^{\pm t/2r_s} \tfrac{\sqrt{\Delta}}{r_s}  \cos \theta \left( \tfrac{r_s^3}{ \Delta} \partial_t \mp \partial_r \Delta \partial_r \mp 2 \tan \theta \partial_\theta \right) , \\ 
	M_{\pm} &= e^{\pm t/2r_s} \cos \varphi  \left[ \tfrac{r_s^2}{\sqrt{\Delta}} \sin \theta \partial_t \mp \tfrac{\sqrt{\Delta} \partial_r \Delta \sin \theta}{r_s } \partial_r \right. \nonumber \\[2pt]
	& \qquad\qquad\qquad\quad  \left. \pm \tfrac{2 \sqrt{\Delta}}{r_s} \cos \theta \partial_\theta \mp \tfrac{2 \sqrt{\Delta}}{r_s} \tfrac{\tan \varphi}{\sin \theta} \partial_\varphi \right] , \\
	N_{\pm} &= e^{\pm t/2r_s} \sin \varphi  \left[ \tfrac{r_s^2}{\sqrt{\Delta}} \sin \theta \partial_t \mp \tfrac{\sqrt{\Delta} \partial_r \Delta \sin \theta}{r_s} \partial_r \right. \nonumber \\[2pt]
	& \qquad\qquad\qquad\quad  \left. \pm \tfrac{2 \sqrt{\Delta}}{r_s} \cos \theta \partial_\theta \pm \tfrac{2 \sqrt{\Delta}}{r_s} \tfrac{\cot \varphi}{\sin \theta} \partial_\varphi \right].
\end{align}
\end{subequations}
Expressing each of the Killing and conformal Killing generators as
$\xi^\mu \partial_\mu$, the symmetries act on the scalar as
\be
\delta \phi = \xi^\mu \partial_\mu\phi +\frac{1}{4}\nabla_\mu\xi^\mu\phi\, .\label{CKV symmetry}
\ee
The time translation $T$ and spatial rotations $J_{ij}$ ($i,j =
1,2,3$) are the familiar symmetries of the exact dynamics. 
In addition,
the symmetry generators $J_{0i}$ ($i=1,2,3$) have a somewhat
privileged status: they generate symmetries of the exact system in
the static limit, $\omega = 0$ \cite{Hui:2021vcv}; see also~\cite{Achour:2022syr} for a related discussion. The other (C)KVs do not give rise to exact symmetries in this limit.
Each contains a factor of $e^{\pm t/2r_s}$, and thus when applied
to a static scalar generates a solution with $\omega=\pm i/2r_s$ (which also means
the resulting scalar has an $|\omega|$ outside the regime of validity of the near-zone
approximation). A corollary is that these other (C)KVs are not well-defined in the
flat space ($r_s \to 0$) limit. Nonetheless, these generators can
still be used to infer properties of exact static
solutions~\cite{Charalambous:2021kcz}. On the other hand, the generators $J_{0i}$ have an overall factor of $1/r_s$ which can be
removed without trouble, and thus do have a
well-defined flat space limit.

\vskip3pt
All together, the algebra of the Killing~\eqref{KVs} and conformal Killing~\eqref{CKVs} symmetries is ${\rm so}(4,2)$, as expected because the metric~\eqref{near zone metric} is conformally flat. There are a number of subalgebras of interest. 
Firstly, the generators $J_{0i} , J_{ij}$ ($i,j = 1,2,3$) form an
${\rm so}(3,1)$ subalgebra~\cite{Hui:2021vcv}. 
In addition, each pair of vectors labeled with the subscripts $\pm$
in eqs.~\eqref{KVs} and \eqref{CKVs} forms a subgroup together with the
generator $T$. More precisely, denoting $X = \{L, K, M,
N\}$, we have 
\begin{equation}
\label{sl2sub}
	[T, X_\pm ] = \pm X_\pm, \qquad \quad [X_+, X_-] = 2 \sigma_X T , 
\end{equation}
with $\sigma_L = -1$ and $\sigma_K= \sigma_M= \sigma_N = +1$, giving different
$\text{\SL}$ subalgebras. To the best of our knowledge, the consequences of the symmetries $K_\pm,
M_\pm$, and $N_\pm$ for perturbations around Schwarzschild have not been
explored in the literature.

\vskip5pt
\noindent
{\bf  Effective Kerr near-zone metric:} The Kerr line element in Boyer--Lindquist coordinates is: 
\be
\begin{aligned}
\dd s^2 
 =\,& -{\rho^2 - r_s r \over \rho^2} \dd t^2  - {2a r_s r \sin^2\theta
  \over \rho^2} \dd t \dd \varphi + {\rho^2 \over \Delta} \dd r^2
  \\
  &+ \rho^2
\dd \theta^2 + { (r^2 + a^2)^2 - a^2\Delta \sin^2\theta \over
  \rho^2} \sin^2\theta \dd \varphi^2
,
\label{Kerrline}
\end{aligned}
\ee
where we have defined the quantities
\be
\rho^2 \equiv r^2 +a^2 \cos^2\theta \, , \qquad
\Delta \equiv r(r-r_s) + a^2 \, .
\ee
The Schwarzschild radius $r_s$ and the spin parameter $a$ are
related to the outer and inner horizons $r_\pm$, i.e.,~the radii where $\Delta = 0$, via $r_\pm \equiv
r_s/2 \pm \sqrt{(r_s/2)^2 - a^2}$.

\vskip3pt
The Klein--Gordon equation on the Kerr background is
\be
\begin{aligned}
\partial_r (\Delta \partial_r \phi) &+ \nabla^2_\Omega \phi- {a^2\over
  \Delta}\partial_\varphi^2 \phi  \\
& - {1\over\Delta} \Big[ (r^2 + a^2)^2 - \Delta a^2 \sin^2\theta\Big] \partial_t^2\phi  \\
& - {2a \over \Delta} \Big[ (r^2 + a^2) - \Delta
  \Big] \partial_t \partial_\varphi \phi = 0 \, .
\end{aligned}
\ee 
We define the near-zone region using the same approximation as in the Schwarzschild case. We choose to implement this approximation in Boyer--Lindquist coordinates because they are inertial at infinity. Wherever
there are time derivatives, we keep terms that go as $1/\Delta$ to preserve
the singularity as $r$ approaches the horizon,
and set $r \rightarrow r_+$ in the corresponding numerators.
This ensures any corrections are subdominant at the horizon and
are small away from it in the low frequency regime: $\omega a \le \omega r_+ \le \omega r \ll 1$. 
Thus the near-zone scalar equation is 
\begin{equation}\label{real space near zone Kerr eq}
\partial_r (\Delta \partial_r \phi) + \nabla^2_\Omega \phi
- {1\over \Delta} \big[ (r_+^2 + a^2) \partial_t + a \partial_\varphi\big]^2
  \phi = 0 \, .
\end{equation} 
Note that, unlike some near-zone approximations of Kerr put forward in the literature, this approximation remains well defined even in the extremal limit $a \to r_s/2$. Expanding in frequency space and spherical harmonics\footnote{For non-static perturbations one should in general use spheroidal harmonics. However, at the order we are working with in the near-zone approximation, it is consistent to  decompose the field in terms of spherical harmonics~\cite{Teukolsky:1973ha}.}
$\phi = e^{-i\omega t} Y_{\ell m} (\theta , \varphi) R(r)$
(where the $\ell$ and $m$ dependence of $R$ is suppressed), 
and using $r_+^2 + a^2 = r_s r_+$, the near-zone equation reads
\cite{1973JETP...37...28S,1974JETP...38....1S,Page:1976df,Maldacena:1997ih}\footnote{Note that there is a typo in eqs.~(2.11) and (2.12) of \cite{Maldacena:1997ih}, where the factor $r_+^4$ should be replaced by $r_s^2r_+^2$.}   
\begin{equation}
\partial_r(\Delta\partial_r R) + \left[ \frac{r_s^2r_+^2}{\Delta}(\omega-m\Omega_+)^2 -\ell(\ell+1)  \right]R=0\, ,
\label{near-zonescalar}
\end{equation}
where for convenience we have introduced $\Omega_+\equiv a/(r_sr_+)$.

\vskip3pt
It is straightforward to show that eq.~\eqref{real space near zone Kerr eq} is the equation of motion for a massless scalar
propagating in the following effective near-zone metric:\footnote{We stress
  that the near-zone metric \eqref{near zone metric Kerr} should not be confused with the near-horizon limit encountered in the context of the extremal Kerr metric (see, e.g.,~\cite{Bardeen:1999px}). This near-horizon limit is defined by a rescaling of the radial and time coordinates which keeps the coordinate $\varphi'$ fixed. As a byproduct, it is suited to study modes with $\omega \sim m \Omega_+$, rather than the static regime.}
\be
\begin{aligned}
\dd s^2_{\text{near-zone}}= \,&- \frac{\Delta - a^2\sin^2\theta}{r_sr_+} \dd t^2 - 2a\sin^2\theta \dd t \dd \varphi 
\\
&+\frac{r_sr_+}{\Delta} \dd r^2+ r_sr_+\dd \Omega^2_{S^2} \, .
\label{near zone metric Kerr}
\end{aligned}
\ee
This metric reduces to~\eqref{near zone
  metric} in the limit $a\to0$, and moreover is conformally flat, and
therefore has the same number of CKVs. A coordinate transformation
$\varphi' = \varphi - (a/r_s r_+) t$ simplifies the metric to
\begin{equation} \label{near zone 2}
\dd s^2_\text{near-zone} = -\frac{\Delta}{\rs\rp}\dd t^2 +
\frac{\rs\rp}{\Delta}\dd r^2+\rs\rp\dd \Omega'_{S^2} {}^2.
\end{equation}
In the extremal limit where $\Delta = (r - r_+)^2 = (r - r_-)^2$, one
can see the $(t, r)$ subspace is ${\rm AdS}_2$ in Poincar\'e
coordinates upon redefining $r-r_-$ as the new radial coordinate. Interestingly, the extremal near-zone metric coincides with
  the \emph{near-horizon} limit of the extremal Reissner--N\"ordstrom
  solution with $r_s^2 \to r_s r_+$.

\vskip3pt
Away from the extremal limit, the simplest way to deduce the
symmetries is to recognize that the effective near-zone metric for
Kerr is in fact equivalent to the near-zone metric for Schwarzschild.
To see this, redefine
$t' = (r_*/\sqrt{r_s r_+}) t$ and $r' = \sqrt{r_s r_+} (r - r_-)/
r_*$, with $r_* \equiv r_+ - r_-$.  The Kerr near-zone metric is then
rewritten as:
\begin{equation} \label{near zone Kerr}
\dd s^2_\text{near-zone} = -\frac{\tilde{\Delta}}{\rs\rp}\dd t'^2 + \frac{\rs\rp}{\tilde{\Delta}}\dd r'^2+\rs\rp \dd\Omega'^2\, ,
\end{equation}
with $\tilde{\Delta} \equiv r'(r'-\sqrt{\rs\rp}) = (r_s r_+/r_*^2)
\Delta$. This has the same form as the Schwarzschild near-zone metric
in~\eqref{near zone metric} with $r_s\to \sqrt{\rs\rp}$. 

\vskip3pt
The 15 (conformal) Killing vectors for a spinning black hole in the
near-zone can thus be obtained from their
Schwarzschild counterparts \eqref{KVs}--\eqref{CKVs} using the
coordinate transformations discussed above and replacing $r_s \to \sqrt{r_s r_+}$. For completeness, we report their explicit expressions in the Supplemental Material. 
There are several \SL{} subalgebras, just as in Schwarzschild, taking the same form
as in eq.~\eqref{sl2sub}. Spatial rotations $J_{ij}$ and
boosts $J_{0i}$ ($i,j = 1,2,3$) form an
${\rm so}(3,1)$ subalgebra just as before. However, only $J_{23}$ among the spatial rotations is an exact symmetry of the
static sector, while $J_{12}$ and $J_{13}$ do not preserve the static nature of field configurations.\footnote{By a static configuration, 
we mean $\partial_t \phi = 0$ keeping $\varphi$ (as well as $r$ and $\theta$)
fixed, as opposed to keeping $\varphi'$ fixed. This choice is dictated by the fact that $\vphi$ is an inertial coordinate at infinity.} Physically, this is because the Kerr metric has a preferred direction. Similarly, of the three boosts,
only $J_{01}$ is a symmetry of the exact system in the static limit.\footnote{This is reflected by the fact that only $J_{01}$ among them has
no time dependence when expressed in Boyer--Lindquist
coordinates.}

\vskip5pt
\noindent
{\bf  Comparison  of different  near zones:} The effective metric~\eqref{near zone metric Kerr}   captures the  near-zone dynamics~\eqref{real space near zone Kerr eq} of  massless scalar perturbations around a Kerr black hole. 
For different reasons, various deformations of the near-zone approximation~\eqref{real space near zone Kerr eq} have been proposed in the literature.
There are in fact many ways of deforming~\eqref{real space near zone
  Kerr eq} at subleading order in 
$(r-r_+)/r_+$~\cite{Teukolsky:1973ha,Lowe:2011aa}. It is instructive to briefly review some of these possibilities and highlight  the main differences with~\eqref{real space near zone Kerr eq}.
 
 \vskip3pt
 One notable example is given in \cite{Castro:2010fd}. Supported by
 the observation that the Cardy formula for a $\text{CFT}_2$ gives
 exactly the Bekenstein--Hawking entropy of the Kerr solution,~\cite{Castro:2010fd} conjectured (see
 also~\cite{Guica:2008mu,Bredberg:2009pv,Bardeen:1999px})  that a
 non-extremal Kerr black hole is dual to a two-dimensional CFT and
 proposed a near-zone approximation with an
 $\text{\SL}_L \times \text{\SL}_R$ symmetry. This can be obtained by
 adding to \eqref{near-zonescalar} the term $(\omega   r_s
 (\omega  r_s^2-2 m a)/(r-r_-))R$. The effect of this term---which is small  in the low frequency regime and subleading near the horizon compared to the $1/\Delta$ term in \eqref{near-zonescalar}---is to break some of the symmetries while introducing a new $\text{\SL}_L \times \text{\SL}_R$ symmetry (which is not a subalgebra of our ${\rm so}(4,2)$). These $\text{\SL}_L \times \text{\SL}_R$ generators are singular in the Schwarzschild limit (the near zone of~\cite{Castro:2010fd} is not smoothly connected to the Schwarzschild near zone above in the limit $a\to0$) and they are not globally defined, as they do not respect the $\varphi\to\varphi+2\pi$ periodicity.
 
 \vskip3pt
A different near-zone approximation that overcomes these issues has recently been proposed in \cite{Charalambous:2021kcz}. The near zone of \cite{Charalambous:2021kcz} differs from the one in eq.~\eqref{real space near zone Kerr eq} by the term $(4\omega\Omega_+ m(r-r_+)/(r_+-r_-) )R$. This too breaks some of the symmetries, but keeps an  $\text{\SL}$  group with generators that are both globally well defined and have a smooth Schwarzschild limit.

\vskip3pt
All three of these approximations are contained in the one-parameter family of~\cite{Lowe:2011aa}, which possesses an $\text{\SL}_L \times \text{\SL}_R$ symmetry except at two special points, one corresponding to~\eqref{real space near zone Kerr eq} (where they identified the generators $T$ \eqref{Tkerr} and $L_\pm$ \eqref{Lkerr}) and the other to~\cite{Charalambous:2021kcz}. See also~\cite{Keeler:2021tqy} for a recent discussion of near-zone approximations and their relations to Killing tensor symmetries.

 \vskip3pt
Each of these approximation schemes has benefits and drawbacks.
For our purposes, the main appeal of the approximation~\eqref{real space near zone Kerr eq}, besides having a smooth Schwarzschild
limit and globally well-defined
generators, is that its effective metric~\eqref{near zone metric Kerr}
contains in particular the symmetry generator  $J_{01}$ in
\eqref{J01kerr}. This is a symmetry of the {\it exact} dynamics for static
field configurations (and is in fact a CKV of an effective 3D
metric~\cite{Hui:2021vcv}). Keeping $J_{01}$ as a symmetry is thus useful
for a near-zone approximation intended for low frequency
phenomena.  
The effective metric \eqref{near zone metric Kerr} has other special
properties: since it is conformally flat, it possesses the maximal
number of CKVs (15 in $d=4$), and since its Ricci scalar vanishes, a
massless scalar is automatically conformally coupled, so that each of
the (C)KVs $\xi^\mu$ generates a symmetry acting on the scalar as
in~\eqref{CKV symmetry}. Each gives rise to a conserved current in the
standard way: $j^\mu = T^{\mu\nu} \xi_\nu$, where $T^{\mu\nu}$ is the (traceless)
energy momentum tensor of the scalar.

\vskip5pt
\noindent
{\bf  Ladder symmetries and tidal response:} 
Up to an irrelevant constant factor, the CKV associated with $J_{01}$
in~\eqref{J01kerr} can be written as $\xi^\mu_{J_{01}} = (0, \Delta
\cos\theta, \frac{1}{2}\Delta' \sin\theta, 0)$, with
$\Delta'\equiv\partial_r\Delta$. As mentioned above, this CKV is time
independent and survives in the zero-frequency limit, recovering the CKV that we showed in~\cite{Hui:2021vcv} to be associated with a ladder structure for static perturbations around  Kerr black holes, leading to vanishing static response.\footnote{Note that the $\text{\SL}$ involving $L_\pm$ in the near zone~\eqref{real space near zone Kerr eq} includes~\eqref{Tkerr} as a generator instead of $\partial_t$, which precludes applying the arguments of~\cite{Charalambous:2021kcz} directly.} Let us now recall how this works and  extend the results of~\cite{Hui:2021vcv} to non-zero frequencies.

\vskip3pt
The CKV $\xi^\mu_{J_{01}}$ corresponds to a symmetry of the scalar action, which acts on the scalar field $\phi$ (in real space) as in~\eqref{CKV symmetry}. After decomposing $\phi$ in spherical harmonics as $\phi = e^{-i \omega t}{Y_{\ell m}}(\theta,\varphi) R_\ell(r)$, and extracting a convenient phase factor from $R_\ell$,
\begin{equation}
R_\ell \equiv e^{-\frac{i r_+ r_s  }{r_+-r_-} \left(\omega -m \Omega_+\right) \log \left(\frac{r-r_+}{r-r_-}\right) } \psi_\ell ,
\end{equation}
the scalar equation \eqref{near-zonescalar} takes the form
\begin{equation}
\left[\partial_r \Big( \Delta  \partial_r  -2 i r_sr_+  \left(\omega -m\Omega_+\right)  \Big)- \ell(\ell+1)\right]\psi_\ell =0 \, ,
\label{near-zonescalarpsi}
\end{equation}
where derivatives act on everything to their right, and the field transformation takes the form
\begin{equation}
\delta \psi_\ell = \mathcal{Q}_\ell D_{\ell-1}^+ \psi_{\ell-1} - \mathcal{Q}_{\ell+1}D_{\ell+1}^- \psi_{\ell+1} \, .
\end{equation}
We have defined $\mathcal{Q}_\ell \equiv \sqrt{(\ell^2-m^2)/(4\ell^2-1)}$
and introduced the operators
\begin{subequations}
\label{Dpmpsi}
\begin{align}
D^+_\ell & \equiv -\Delta \partial_r - \frac{\ell+1}{2}\Delta'  + i r_s r_+(\omega-m \Omega_+) ,
\label{Dppsi}
\\
D^-_\ell & \equiv \Delta \partial_r - \frac{\ell}{2}\Delta' -  i r_s r_+(\omega-m \Omega_+).
\label{Dmpsi}
\end{align}
\end{subequations}
The $D^\pm_\ell$ are \textit{ladder operators} in the sense that $\delta \psi_{\ell\pm1}=D_\ell^{\pm}\psi_\ell$ are  solutions to the near-zone equation \eqref{near-zonescalarpsi} at level $\ell\pm1$ if $\psi_\ell$ solves \eqref{near-zonescalarpsi} at level $\ell$. These operators are useful because they allow us to recursively define an on-shell conserved charge  at each $\ell$ \cite{Hui:2021vcv}:
\begin{equation}
P_\ell  = \alpha_\ell \big[  \Delta  \partial_r  -2 i r_sr_+  \left(\omega -m\Omega_+\right) \big] D_1^- \cdots D_\ell^- \psi_\ell ,
\end{equation}
where $\alpha_\ell\equiv
-2^{2\ell-1}(\ell!)^2/[(2\ell)!(2\ell+1)!]$. These conserved
  charges allow us to connect the behavior of solutions near the
  horizon to the behavior at infinity without solving eq.~\eqref{near-zonescalarpsi} explicitly, making it possible
  to infer the vanishing of static responses from
  symmetry~\cite{Hui:2021vcv}. Corresponding to each of these charges
is an off-shell symmetry of the action, which can be inferred
as described in \cite{Hui:2021vcv} (see also~\cite{Compton:2020cjx} for a similar construction on de Sitter backgrounds).

\vskip3pt
Evaluating $P_\ell$ on $\psi_\ell= D_{\ell-1}^+\cdots D_0^+\psi_0$, where $\psi_0={\rm constant}$---which is the solution with the correct infalling behavior at $r=r_+$---yields
\begin{equation}
P_\ell  =  -iq (r_+-r_-)^{2\ell+1} \frac{(\ell!)^2}{(2\ell)!(2\ell+1)!} \prod_{k=1}^\ell (k^2+4q^2) ,
\label{growingPl}
\end{equation}
with
$q\equiv r_sr_+(m\Omega_+-\omega)/(r_+-r_-)$.
The conserved charges $P_\ell$ in \eqref{growingPl} reproduce
the induced multipole moments of a  scalar field on a fixed Kerr geometry~\cite{Wong:2019yoc} (see also~\cite{1973JETP...37...28S,1974JETP...38....1S,Creci:2021rkz}). In particular, when $\omega=0$, one recovers the scalar's static dissipative response~\cite{LeTiec:2020spy,LeTiec:2020bos,Chia:2020yla,Charalambous:2021mea,Hui:2021vcv}.\\


\noindent 
{\bf Acknowledgments:}  We would like to thank Albert Law and John Stout for helpful discussions. LH is supported by the DOE DE-SC0011941 and a Simons Fellowship in Theoretical Physics. The work of RP is supported in part by the National Science Foundation under Grant No. PHY-1915611. ARS's research was partially supported by funds from the Natural Sciences and Engineering Research Council (NSERC) of Canada. Research at the Perimeter Institute is supported in part by the Government of Canada through NSERC and by the Province of Ontario through MRI.

\newpage
\bibliographystyle{apsrev4-1}
\bibliography{biblio}

\begin{thebibliography}{33}%
\makeatletter
\providecommand \@ifxundefined [1]{%
 \@ifx{#1\undefined}
}%
\providecommand \@ifnum [1]{%
 \ifnum #1\expandafter \@firstoftwo
 \else \expandafter \@secondoftwo
 \fi
}%
\providecommand \@ifx [1]{%
 \ifx #1\expandafter \@firstoftwo
 \else \expandafter \@secondoftwo
 \fi
}%
\providecommand \natexlab [1]{#1}%
\providecommand \enquote  [1]{``#1''}%
\providecommand \bibnamefont  [1]{#1}%
\providecommand \bibfnamefont [1]{#1}%
\providecommand \citenamefont [1]{#1}%
\providecommand \href@noop [0]{\@secondoftwo}%
\providecommand \href [0]{\begingroup \@sanitize@url \@href}%
\providecommand \@href[1]{\@@startlink{#1}\@@href}%
\providecommand \@@href[1]{\endgroup#1\@@endlink}%
\providecommand \@sanitize@url [0]{\catcode `\\12\catcode `\$12\catcode
  `\&12\catcode `\#12\catcode `\^12\catcode `\_12\catcode `\%12\relax}%
\providecommand \@@startlink[1]{}%
\providecommand \@@endlink[0]{}%
\providecommand \url  [0]{\begingroup\@sanitize@url \@url }%
\providecommand \@url [1]{\endgroup\@href {#1}{\urlprefix }}%
\providecommand \urlprefix  [0]{URL }%
\providecommand \Eprint [0]{\href }%
\providecommand \doibase [0]{http://dx.doi.org/}%
\providecommand \selectlanguage [0]{\@gobble}%
\providecommand \bibinfo  [0]{\@secondoftwo}%
\providecommand \bibfield  [0]{\@secondoftwo}%
\providecommand \translation [1]{[#1]}%
\providecommand \BibitemOpen [0]{}%
\providecommand \bibitemStop [0]{}%
\providecommand \bibitemNoStop [0]{.\EOS\space}%
\providecommand \EOS [0]{\spacefactor3000\relax}%
\providecommand \BibitemShut  [1]{\csname bibitem#1\endcsname}%
\let\auto@bib@innerbib\@empty
\bibitem [{\citenamefont {Regge}\ and\ \citenamefont
  {Wheeler}(1957)}]{Regge:1957td}%
  \BibitemOpen
  \bibfield  {author} {\bibinfo {author} {\bibfnamefont {T.}~\bibnamefont
  {Regge}}\ and\ \bibinfo {author} {\bibfnamefont {J.~A.}\ \bibnamefont
  {Wheeler}},\ }\href {\doibase 10.1103/PhysRev.108.1063} {\bibfield  {journal}
  {\bibinfo  {journal} {Phys. Rev.}\ }\textbf {\bibinfo {volume} {108}},\
  \bibinfo {pages} {1063} (\bibinfo {year} {1957})}\BibitemShut {NoStop}%
\bibitem [{\citenamefont {Zerilli}(1970)}]{Zerilli:1970se}%
  \BibitemOpen
  \bibfield  {author} {\bibinfo {author} {\bibfnamefont {F.~J.}\ \bibnamefont
  {Zerilli}},\ }\href {\doibase 10.1103/PhysRevLett.24.737} {\bibfield
  {journal} {\bibinfo  {journal} {Phys. Rev. Lett.}\ }\textbf {\bibinfo
  {volume} {24}},\ \bibinfo {pages} {737} (\bibinfo {year} {1970})}\BibitemShut
  {NoStop}%
\bibitem [{\citenamefont {Charalambous}\ \emph
  {et~al.}(2021{\natexlab{a}})\citenamefont {Charalambous}, \citenamefont
  {Dubovsky},\ and\ \citenamefont {Ivanov}}]{Charalambous:2021kcz}%
  \BibitemOpen
  \bibfield  {author} {\bibinfo {author} {\bibfnamefont {P.}~\bibnamefont
  {Charalambous}}, \bibinfo {author} {\bibfnamefont {S.}~\bibnamefont
  {Dubovsky}}, \ and\ \bibinfo {author} {\bibfnamefont {M.~M.}\ \bibnamefont
  {Ivanov}},\ }\href {\doibase 10.1103/PhysRevLett.127.101101} {\bibfield
  {journal} {\bibinfo  {journal} {Phys. Rev. Lett.}\ }\textbf {\bibinfo
  {volume} {127}},\ \bibinfo {pages} {101101} (\bibinfo {year}
  {2021}{\natexlab{a}})},\ \Eprint {http://arxiv.org/abs/2103.01234}
  {arXiv:2103.01234 [hep-th]} \BibitemShut {NoStop}%
\bibitem [{\citenamefont {Hui}\ \emph {et~al.}(2022)\citenamefont {Hui},
  \citenamefont {Joyce}, \citenamefont {Penco}, \citenamefont {Santoni},\ and\
  \citenamefont {Solomon}}]{Hui:2021vcv}%
  \BibitemOpen
  \bibfield  {author} {\bibinfo {author} {\bibfnamefont {L.}~\bibnamefont
  {Hui}}, \bibinfo {author} {\bibfnamefont {A.}~\bibnamefont {Joyce}}, \bibinfo
  {author} {\bibfnamefont {R.}~\bibnamefont {Penco}}, \bibinfo {author}
  {\bibfnamefont {L.}~\bibnamefont {Santoni}}, \ and\ \bibinfo {author}
  {\bibfnamefont {A.~R.}\ \bibnamefont {Solomon}},\ }\href {\doibase
  10.1088/1475-7516/2022/01/032} {\bibfield  {journal} {\bibinfo  {journal}
  {JCAP}\ }\textbf {\bibinfo {volume} {01}},\ \bibinfo {pages} {032} (\bibinfo
  {year} {2022})},\ \Eprint {http://arxiv.org/abs/2105.01069} {arXiv:2105.01069
  [hep-th]} \BibitemShut {NoStop}%
\bibitem [{\citenamefont {Fang}\ and\ \citenamefont
  {Lovelace}(2005)}]{Fang:2005qq}%
  \BibitemOpen
  \bibfield  {author} {\bibinfo {author} {\bibfnamefont {H.}~\bibnamefont
  {Fang}}\ and\ \bibinfo {author} {\bibfnamefont {G.}~\bibnamefont
  {Lovelace}},\ }\href {\doibase 10.1103/PhysRevD.72.124016} {\bibfield
  {journal} {\bibinfo  {journal} {Phys. Rev. D}\ }\textbf {\bibinfo {volume}
  {72}},\ \bibinfo {pages} {124016} (\bibinfo {year} {2005})},\ \Eprint
  {http://arxiv.org/abs/gr-qc/0505156} {arXiv:gr-qc/0505156} \BibitemShut
  {NoStop}%
\bibitem [{\citenamefont {Damour}\ and\ \citenamefont
  {Nagar}(2009)}]{Damour:2009vw}%
  \BibitemOpen
  \bibfield  {author} {\bibinfo {author} {\bibfnamefont {T.}~\bibnamefont
  {Damour}}\ and\ \bibinfo {author} {\bibfnamefont {A.}~\bibnamefont {Nagar}},\
  }\href {\doibase 10.1103/PhysRevD.80.084035} {\bibfield  {journal} {\bibinfo
  {journal} {Phys. Rev. D}\ }\textbf {\bibinfo {volume} {80}},\ \bibinfo
  {pages} {084035} (\bibinfo {year} {2009})},\ \Eprint
  {http://arxiv.org/abs/0906.0096} {arXiv:0906.0096 [gr-qc]} \BibitemShut
  {NoStop}%
\bibitem [{\citenamefont {Binnington}\ and\ \citenamefont
  {Poisson}(2009)}]{Binnington:2009bb}%
  \BibitemOpen
  \bibfield  {author} {\bibinfo {author} {\bibfnamefont {T.}~\bibnamefont
  {Binnington}}\ and\ \bibinfo {author} {\bibfnamefont {E.}~\bibnamefont
  {Poisson}},\ }\href {\doibase 10.1103/PhysRevD.80.084018} {\bibfield
  {journal} {\bibinfo  {journal} {Phys. Rev.}\ }\textbf {\bibinfo {volume}
  {D80}},\ \bibinfo {pages} {084018} (\bibinfo {year} {2009})},\ \Eprint
  {http://arxiv.org/abs/0906.1366} {arXiv:0906.1366 [gr-qc]} \BibitemShut
  {NoStop}%
\bibitem [{\citenamefont {Damour}\ and\ \citenamefont
  {Lecian}(2009)}]{Damour:2009va}%
  \BibitemOpen
  \bibfield  {author} {\bibinfo {author} {\bibfnamefont {T.}~\bibnamefont
  {Damour}}\ and\ \bibinfo {author} {\bibfnamefont {O.~M.}\ \bibnamefont
  {Lecian}},\ }\href {\doibase 10.1103/PhysRevD.80.044017} {\bibfield
  {journal} {\bibinfo  {journal} {Phys. Rev.}\ }\textbf {\bibinfo {volume}
  {D80}},\ \bibinfo {pages} {044017} (\bibinfo {year} {2009})},\ \Eprint
  {http://arxiv.org/abs/0906.3003} {arXiv:0906.3003 [gr-qc]} \BibitemShut
  {NoStop}%
\bibitem [{\citenamefont {Kol}\ and\ \citenamefont
  {Smolkin}(2012)}]{Kol:2011vg}%
  \BibitemOpen
  \bibfield  {author} {\bibinfo {author} {\bibfnamefont {B.}~\bibnamefont
  {Kol}}\ and\ \bibinfo {author} {\bibfnamefont {M.}~\bibnamefont {Smolkin}},\
  }\href@noop {} {\bibfield  {journal} {\bibinfo  {journal} {JHEP}\ }\textbf
  {\bibinfo {volume} {02}},\ \bibinfo {pages} {010} (\bibinfo {year} {2012})},\
  \Eprint {http://arxiv.org/abs/1110.3764} {arXiv:1110.3764 [hep-th]}
  \BibitemShut {NoStop}%
\bibitem [{\citenamefont {Le~Tiec}\ and\ \citenamefont
  {Casals}(2021)}]{LeTiec:2020spy}%
  \BibitemOpen
  \bibfield  {author} {\bibinfo {author} {\bibfnamefont {A.}~\bibnamefont
  {Le~Tiec}}\ and\ \bibinfo {author} {\bibfnamefont {M.}~\bibnamefont
  {Casals}},\ }\href {\doibase 10.1103/PhysRevLett.126.131102} {\bibfield
  {journal} {\bibinfo  {journal} {Phys. Rev. Lett.}\ }\textbf {\bibinfo
  {volume} {126}},\ \bibinfo {pages} {131102} (\bibinfo {year} {2021})},\
  \Eprint {http://arxiv.org/abs/2007.00214} {arXiv:2007.00214 [gr-qc]}
  \BibitemShut {NoStop}%
\bibitem [{\citenamefont {Le~Tiec}\ \emph {et~al.}(2021)\citenamefont
  {Le~Tiec}, \citenamefont {Casals},\ and\ \citenamefont
  {Franzin}}]{LeTiec:2020bos}%
  \BibitemOpen
  \bibfield  {author} {\bibinfo {author} {\bibfnamefont {A.}~\bibnamefont
  {Le~Tiec}}, \bibinfo {author} {\bibfnamefont {M.}~\bibnamefont {Casals}}, \
  and\ \bibinfo {author} {\bibfnamefont {E.}~\bibnamefont {Franzin}},\ }\href
  {\doibase 10.1103/PhysRevD.103.084021} {\bibfield  {journal} {\bibinfo
  {journal} {Phys. Rev. D}\ }\textbf {\bibinfo {volume} {103}},\ \bibinfo
  {pages} {084021} (\bibinfo {year} {2021})},\ \Eprint
  {http://arxiv.org/abs/2010.15795} {arXiv:2010.15795 [gr-qc]} \BibitemShut
  {NoStop}%
\bibitem [{\citenamefont {Chia}(2021)}]{Chia:2020yla}%
  \BibitemOpen
  \bibfield  {author} {\bibinfo {author} {\bibfnamefont {H.~S.}\ \bibnamefont
  {Chia}},\ }\href {\doibase 10.1103/PhysRevD.104.024013} {\bibfield  {journal}
  {\bibinfo  {journal} {Phys. Rev. D}\ }\textbf {\bibinfo {volume} {104}},\
  \bibinfo {pages} {024013} (\bibinfo {year} {2021})},\ \Eprint
  {http://arxiv.org/abs/2010.07300} {arXiv:2010.07300 [gr-qc]} \BibitemShut
  {NoStop}%
\bibitem [{\citenamefont {Hui}\ \emph {et~al.}(2021)\citenamefont {Hui},
  \citenamefont {Joyce}, \citenamefont {Penco}, \citenamefont {Santoni},\ and\
  \citenamefont {Solomon}}]{Hui:2020xxx}%
  \BibitemOpen
  \bibfield  {author} {\bibinfo {author} {\bibfnamefont {L.}~\bibnamefont
  {Hui}}, \bibinfo {author} {\bibfnamefont {A.}~\bibnamefont {Joyce}}, \bibinfo
  {author} {\bibfnamefont {R.}~\bibnamefont {Penco}}, \bibinfo {author}
  {\bibfnamefont {L.}~\bibnamefont {Santoni}}, \ and\ \bibinfo {author}
  {\bibfnamefont {A.~R.}\ \bibnamefont {Solomon}},\ }\href {\doibase
  10.1088/1475-7516/2021/04/052} {\bibfield  {journal} {\bibinfo  {journal}
  {JCAP}\ }\textbf {\bibinfo {volume} {04}},\ \bibinfo {pages} {052} (\bibinfo
  {year} {2021})},\ \Eprint {http://arxiv.org/abs/2010.00593} {arXiv:2010.00593
  [hep-th]} \BibitemShut {NoStop}%
\bibitem [{\citenamefont {Charalambous}\ \emph
  {et~al.}(2021{\natexlab{b}})\citenamefont {Charalambous}, \citenamefont
  {Dubovsky},\ and\ \citenamefont {Ivanov}}]{Charalambous:2021mea}%
  \BibitemOpen
  \bibfield  {author} {\bibinfo {author} {\bibfnamefont {P.}~\bibnamefont
  {Charalambous}}, \bibinfo {author} {\bibfnamefont {S.}~\bibnamefont
  {Dubovsky}}, \ and\ \bibinfo {author} {\bibfnamefont {M.~M.}\ \bibnamefont
  {Ivanov}},\ }\href {\doibase 10.1007/JHEP05(2021)038} {\bibfield  {journal}
  {\bibinfo  {journal} {JHEP}\ }\textbf {\bibinfo {volume} {05}},\ \bibinfo
  {pages} {038} (\bibinfo {year} {2021}{\natexlab{b}})},\ \Eprint
  {http://arxiv.org/abs/2102.08917} {arXiv:2102.08917 [hep-th]} \BibitemShut
  {NoStop}%
\bibitem [{\citenamefont {{Starobinski{\v{i}}}}(1973)}]{1973JETP...37...28S}%
  \BibitemOpen
  \bibfield  {author} {\bibinfo {author} {\bibfnamefont {A.~A.}\ \bibnamefont
  {{Starobinski{\v{i}}}}},\ }\href@noop {} {\bibfield  {journal} {\bibinfo
  {journal} {Soviet Journal of Experimental and Theoretical Physics}\ }\textbf
  {\bibinfo {volume} {37}},\ \bibinfo {pages} {28} (\bibinfo {year}
  {1973})}\BibitemShut {NoStop}%
\bibitem [{\citenamefont {{Starobinski{\v{i}}}}\ and\ \citenamefont
  {{Churilov}}(1974)}]{1974JETP...38....1S}%
  \BibitemOpen
  \bibfield  {author} {\bibinfo {author} {\bibfnamefont {A.~A.}\ \bibnamefont
  {{Starobinski{\v{i}}}}}\ and\ \bibinfo {author} {\bibfnamefont {S.~M.}\
  \bibnamefont {{Churilov}}},\ }\href@noop {} {\bibfield  {journal} {\bibinfo
  {journal} {Soviet Journal of Experimental and Theoretical Physics}\ }\textbf
  {\bibinfo {volume} {38}},\ \bibinfo {pages} {1} (\bibinfo {year}
  {1974})}\BibitemShut {NoStop}%
\bibitem [{\citenamefont {Bertini}\ \emph {et~al.}(2012)\citenamefont
  {Bertini}, \citenamefont {Cacciatori},\ and\ \citenamefont
  {Klemm}}]{Bertini:2011ga}%
  \BibitemOpen
  \bibfield  {author} {\bibinfo {author} {\bibfnamefont {S.}~\bibnamefont
  {Bertini}}, \bibinfo {author} {\bibfnamefont {S.~L.}\ \bibnamefont
  {Cacciatori}}, \ and\ \bibinfo {author} {\bibfnamefont {D.}~\bibnamefont
  {Klemm}},\ }\href {\doibase 10.1103/PhysRevD.85.064018} {\bibfield  {journal}
  {\bibinfo  {journal} {Phys. Rev. D}\ }\textbf {\bibinfo {volume} {85}},\
  \bibinfo {pages} {064018} (\bibinfo {year} {2012})},\ \Eprint
  {http://arxiv.org/abs/1106.0999} {arXiv:1106.0999 [hep-th]} \BibitemShut
  {NoStop}%
\bibitem [{\citenamefont {Cvetic}\ \emph {et~al.}(2021)\citenamefont {Cvetic},
  \citenamefont {Gibbons}, \citenamefont {Pope},\ and\ \citenamefont
  {Whiting}}]{Cvetic:2021vxa}%
  \BibitemOpen
  \bibfield  {author} {\bibinfo {author} {\bibfnamefont {M.}~\bibnamefont
  {Cvetic}}, \bibinfo {author} {\bibfnamefont {G.~W.}\ \bibnamefont {Gibbons}},
  \bibinfo {author} {\bibfnamefont {C.~N.}\ \bibnamefont {Pope}}, \ and\
  \bibinfo {author} {\bibfnamefont {B.~F.}\ \bibnamefont {Whiting}},\
  }\href@noop {} {\  (\bibinfo {year} {2021})},\ \Eprint
  {http://arxiv.org/abs/2109.03254} {arXiv:2109.03254 [gr-qc]} \BibitemShut
  {NoStop}%
\bibitem [{\citenamefont {Achour}\ \emph {et~al.}(2022)\citenamefont {Achour},
  \citenamefont {Livine}, \citenamefont {Mukohyama},\ and\ \citenamefont
  {Uzan}}]{Achour:2022syr}%
  \BibitemOpen
  \bibfield  {author} {\bibinfo {author} {\bibfnamefont {J.~B.}\ \bibnamefont
  {Achour}}, \bibinfo {author} {\bibfnamefont {E.~R.}\ \bibnamefont {Livine}},
  \bibinfo {author} {\bibfnamefont {S.}~\bibnamefont {Mukohyama}}, \ and\
  \bibinfo {author} {\bibfnamefont {J.-P.}\ \bibnamefont {Uzan}},\ }\href@noop
  {} {\  (\bibinfo {year} {2022})},\ \Eprint {http://arxiv.org/abs/2202.12828}
  {arXiv:2202.12828 [gr-qc]} \BibitemShut {NoStop}%
\bibitem [{\citenamefont {Teukolsky}(1973)}]{Teukolsky:1973ha}%
  \BibitemOpen
  \bibfield  {author} {\bibinfo {author} {\bibfnamefont {S.~A.}\ \bibnamefont
  {Teukolsky}},\ }\href {\doibase 10.1086/152444} {\bibfield  {journal}
  {\bibinfo  {journal} {Astrophys. J.}\ }\textbf {\bibinfo {volume} {185}},\
  \bibinfo {pages} {635} (\bibinfo {year} {1973})}\BibitemShut {NoStop}%
\bibitem [{\citenamefont {Page}(1976)}]{Page:1976df}%
  \BibitemOpen
  \bibfield  {author} {\bibinfo {author} {\bibfnamefont {D.~N.}\ \bibnamefont
  {Page}},\ }\href {\doibase 10.1103/PhysRevD.13.198} {\bibfield  {journal}
  {\bibinfo  {journal} {Phys. Rev. D}\ }\textbf {\bibinfo {volume} {13}},\
  \bibinfo {pages} {198} (\bibinfo {year} {1976})}\BibitemShut {NoStop}%
\bibitem [{\citenamefont {Maldacena}\ and\ \citenamefont
  {Strominger}(1997)}]{Maldacena:1997ih}%
  \BibitemOpen
  \bibfield  {author} {\bibinfo {author} {\bibfnamefont {J.~M.}\ \bibnamefont
  {Maldacena}}\ and\ \bibinfo {author} {\bibfnamefont {A.}~\bibnamefont
  {Strominger}},\ }\href {\doibase 10.1103/PhysRevD.56.4975} {\bibfield
  {journal} {\bibinfo  {journal} {Phys. Rev. D}\ }\textbf {\bibinfo {volume}
  {56}},\ \bibinfo {pages} {4975} (\bibinfo {year} {1997})},\ \Eprint
  {http://arxiv.org/abs/hep-th/9702015} {arXiv:hep-th/9702015} \BibitemShut
  {NoStop}%
\bibitem [{\citenamefont {Bardeen}\ and\ \citenamefont
  {Horowitz}(1999)}]{Bardeen:1999px}%
  \BibitemOpen
  \bibfield  {author} {\bibinfo {author} {\bibfnamefont {J.~M.}\ \bibnamefont
  {Bardeen}}\ and\ \bibinfo {author} {\bibfnamefont {G.~T.}\ \bibnamefont
  {Horowitz}},\ }\href {\doibase 10.1103/PhysRevD.60.104030} {\bibfield
  {journal} {\bibinfo  {journal} {Phys. Rev. D}\ }\textbf {\bibinfo {volume}
  {60}},\ \bibinfo {pages} {104030} (\bibinfo {year} {1999})},\ \Eprint
  {http://arxiv.org/abs/hep-th/9905099} {arXiv:hep-th/9905099} \BibitemShut
  {NoStop}%
\bibitem [{\citenamefont {Lowe}\ and\ \citenamefont
  {Skanata}(2012)}]{Lowe:2011aa}%
  \BibitemOpen
  \bibfield  {author} {\bibinfo {author} {\bibfnamefont {D.~A.}\ \bibnamefont
  {Lowe}}\ and\ \bibinfo {author} {\bibfnamefont {A.}~\bibnamefont {Skanata}},\
  }\href {\doibase 10.1088/1751-8113/45/47/475401} {\bibfield  {journal}
  {\bibinfo  {journal} {J. Phys. A}\ }\textbf {\bibinfo {volume} {45}},\
  \bibinfo {pages} {475401} (\bibinfo {year} {2012})},\ \Eprint
  {http://arxiv.org/abs/1112.1431} {arXiv:1112.1431 [hep-th]} \BibitemShut
  {NoStop}%
\bibitem [{\citenamefont {Castro}\ \emph {et~al.}(2010)\citenamefont {Castro},
  \citenamefont {Maloney},\ and\ \citenamefont {Strominger}}]{Castro:2010fd}%
  \BibitemOpen
  \bibfield  {author} {\bibinfo {author} {\bibfnamefont {A.}~\bibnamefont
  {Castro}}, \bibinfo {author} {\bibfnamefont {A.}~\bibnamefont {Maloney}}, \
  and\ \bibinfo {author} {\bibfnamefont {A.}~\bibnamefont {Strominger}},\
  }\href {\doibase 10.1103/PhysRevD.82.024008} {\bibfield  {journal} {\bibinfo
  {journal} {Phys. Rev. D}\ }\textbf {\bibinfo {volume} {82}},\ \bibinfo
  {pages} {024008} (\bibinfo {year} {2010})},\ \Eprint
  {http://arxiv.org/abs/1004.0996} {arXiv:1004.0996 [hep-th]} \BibitemShut
  {NoStop}%
\bibitem [{\citenamefont {Guica}\ \emph {et~al.}(2009)\citenamefont {Guica},
  \citenamefont {Hartman}, \citenamefont {Song},\ and\ \citenamefont
  {Strominger}}]{Guica:2008mu}%
  \BibitemOpen
  \bibfield  {author} {\bibinfo {author} {\bibfnamefont {M.}~\bibnamefont
  {Guica}}, \bibinfo {author} {\bibfnamefont {T.}~\bibnamefont {Hartman}},
  \bibinfo {author} {\bibfnamefont {W.}~\bibnamefont {Song}}, \ and\ \bibinfo
  {author} {\bibfnamefont {A.}~\bibnamefont {Strominger}},\ }\href {\doibase
  10.1103/PhysRevD.80.124008} {\bibfield  {journal} {\bibinfo  {journal} {Phys.
  Rev. D}\ }\textbf {\bibinfo {volume} {80}},\ \bibinfo {pages} {124008}
  (\bibinfo {year} {2009})},\ \Eprint {http://arxiv.org/abs/0809.4266}
  {arXiv:0809.4266 [hep-th]} \BibitemShut {NoStop}%
\bibitem [{\citenamefont {Bredberg}\ \emph {et~al.}(2010)\citenamefont
  {Bredberg}, \citenamefont {Hartman}, \citenamefont {Song},\ and\
  \citenamefont {Strominger}}]{Bredberg:2009pv}%
  \BibitemOpen
  \bibfield  {author} {\bibinfo {author} {\bibfnamefont {I.}~\bibnamefont
  {Bredberg}}, \bibinfo {author} {\bibfnamefont {T.}~\bibnamefont {Hartman}},
  \bibinfo {author} {\bibfnamefont {W.}~\bibnamefont {Song}}, \ and\ \bibinfo
  {author} {\bibfnamefont {A.}~\bibnamefont {Strominger}},\ }\href {\doibase
  10.1007/JHEP04(2010)019} {\bibfield  {journal} {\bibinfo  {journal} {JHEP}\
  }\textbf {\bibinfo {volume} {04}},\ \bibinfo {pages} {019} (\bibinfo {year}
  {2010})},\ \Eprint {http://arxiv.org/abs/0907.3477} {arXiv:0907.3477
  [hep-th]} \BibitemShut {NoStop}%
\bibitem [{\citenamefont {Keeler}\ \emph {et~al.}(2021)\citenamefont {Keeler},
  \citenamefont {Martin},\ and\ \citenamefont {Priya}}]{Keeler:2021tqy}%
  \BibitemOpen
  \bibfield  {author} {\bibinfo {author} {\bibfnamefont {C.}~\bibnamefont
  {Keeler}}, \bibinfo {author} {\bibfnamefont {V.}~\bibnamefont {Martin}}, \
  and\ \bibinfo {author} {\bibfnamefont {A.}~\bibnamefont {Priya}},\
  }\href@noop {} {\  (\bibinfo {year} {2021})},\ \Eprint
  {http://arxiv.org/abs/2110.10723} {arXiv:2110.10723 [hep-th]} \BibitemShut
  {NoStop}%
\bibitem [{\citenamefont {Compton}\ and\ \citenamefont
  {Morrison}(2020)}]{Compton:2020cjx}%
  \BibitemOpen
  \bibfield  {author} {\bibinfo {author} {\bibfnamefont {G.}~\bibnamefont
  {Compton}}\ and\ \bibinfo {author} {\bibfnamefont {I.~A.}\ \bibnamefont
  {Morrison}},\ }\href@noop {} {\bibfield  {journal} {\bibinfo  {journal}
  {Class. Quant. Grav.}\ }\textbf {\bibinfo {volume} {37}},\ \bibinfo {pages}
  {125001} (\bibinfo {year} {2020})},\ \Eprint
  {http://arxiv.org/abs/2003.08023} {arXiv:2003.08023 [gr-qc]} \BibitemShut
  {NoStop}%
\bibitem [{\citenamefont {Wong}\ \emph {et~al.}(2019)\citenamefont {Wong},
  \citenamefont {Davis},\ and\ \citenamefont {Gregory}}]{Wong:2019yoc}%
  \BibitemOpen
  \bibfield  {author} {\bibinfo {author} {\bibfnamefont {L.~K.}\ \bibnamefont
  {Wong}}, \bibinfo {author} {\bibfnamefont {A.-C.}\ \bibnamefont {Davis}}, \
  and\ \bibinfo {author} {\bibfnamefont {R.}~\bibnamefont {Gregory}},\ }\href
  {\doibase 10.1103/PhysRevD.100.024010} {\bibfield  {journal} {\bibinfo
  {journal} {Phys. Rev. D}\ }\textbf {\bibinfo {volume} {100}},\ \bibinfo
  {pages} {024010} (\bibinfo {year} {2019})},\ \Eprint
  {http://arxiv.org/abs/1903.07080} {arXiv:1903.07080 [hep-th]} \BibitemShut
  {NoStop}%
\bibitem [{\citenamefont {Creci}\ \emph {et~al.}(2021)\citenamefont {Creci},
  \citenamefont {Hinderer},\ and\ \citenamefont {Steinhoff}}]{Creci:2021rkz}%
  \BibitemOpen
  \bibfield  {author} {\bibinfo {author} {\bibfnamefont {G.}~\bibnamefont
  {Creci}}, \bibinfo {author} {\bibfnamefont {T.}~\bibnamefont {Hinderer}}, \
  and\ \bibinfo {author} {\bibfnamefont {J.}~\bibnamefont {Steinhoff}},\ }\href
  {\doibase 10.1103/PhysRevD.104.124061} {\bibfield  {journal} {\bibinfo
  {journal} {Phys. Rev. D}\ }\textbf {\bibinfo {volume} {104}},\ \bibinfo
  {pages} {124061} (\bibinfo {year} {2021})},\ \Eprint
  {http://arxiv.org/abs/2108.03385} {arXiv:2108.03385 [gr-qc]} \BibitemShut
  {NoStop}%
\bibitem [{\citenamefont {Press}\ and\ \citenamefont
  {Teukolsky}(1973)}]{Press:1973zz}%
  \BibitemOpen
  \bibfield  {author} {\bibinfo {author} {\bibfnamefont {W.~H.}\ \bibnamefont
  {Press}}\ and\ \bibinfo {author} {\bibfnamefont {S.~A.}\ \bibnamefont
  {Teukolsky}},\ }\href {\doibase 10.1086/152445} {\bibfield  {journal}
  {\bibinfo  {journal} {Astrophys. J.}\ }\textbf {\bibinfo {volume} {185}},\
  \bibinfo {pages} {649} (\bibinfo {year} {1973})}\BibitemShut {NoStop}%
\bibitem [{\citenamefont {Chandrasekhar}(1985)}]{Chandrasekhar:1985kt}%
  \BibitemOpen
  \bibfield  {author} {\bibinfo {author} {\bibfnamefont {S.}~\bibnamefont
  {Chandrasekhar}},\ }\href@noop {} {\emph {\bibinfo {title} {{The mathematical
  theory of black holes}}}}\ (\bibinfo {year} {1985})\BibitemShut {NoStop}%
\end{thebibliography}%

\onecolumngrid

\newpage

\onecolumngrid

\section*{Supplemental material}

\subsection{Near-zone AdS$_2$ geometry}
\begin{figure}[h!] 
	\includegraphics[scale=.8]{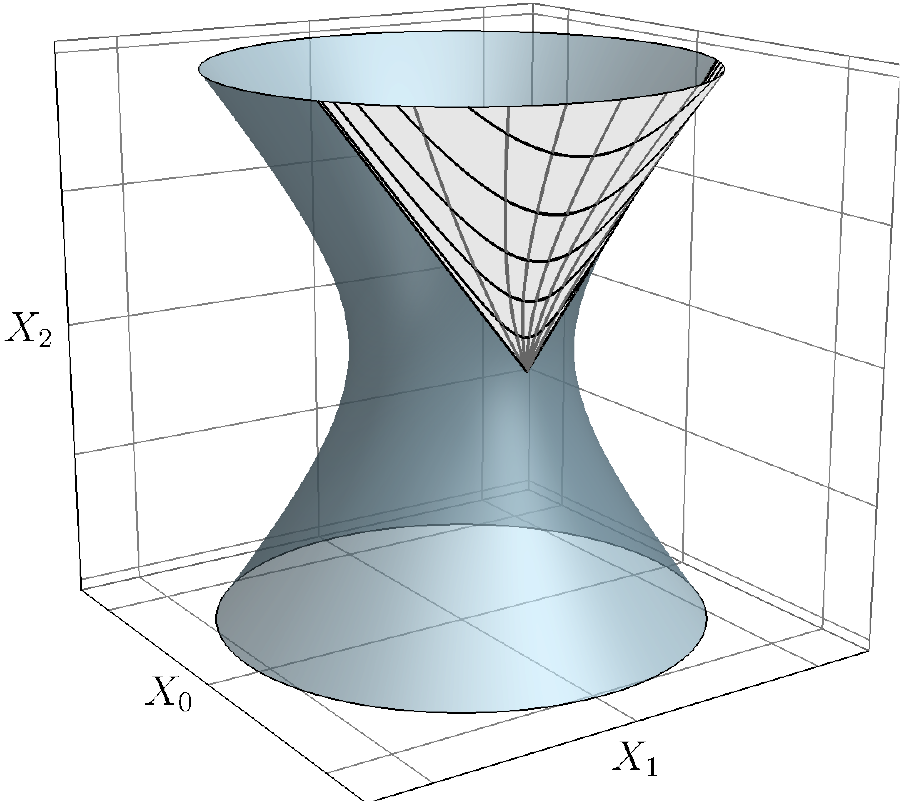}
	\caption{\small Portion of ${\rm AdS}_2$ covered by the de Sitter slice coordinates $(\tau, \xi)$, along with lines of constant $\tau$ and $\xi$.}\label{fig} 
\end{figure}

\noindent
Here we elaborate on the geometry of the 
$(t,r)$ subspace of the near-zone metric~\eqref{near zone metric}. This metric describes ${\rm AdS}_2$ in de Sitter-slice coordinates. In order to see this, we make the coordinate transformation
\begin{equation}\label{AdS coords}
	\tau = \frac{t}{2 r_s} \, ,
	\qquad\quad
	\xi = \cosh^{-1} \left(\frac{2 r}{r_s} -1\right) \, ,
\end{equation}
so that the two-dimensional metric $\dd s^2= - \frac{\Delta}{r_s^2} \dd t^2 + \frac{r_s^2 }{\Delta} \dd r^2$ becomes
\begin{align}
\label{AdS2 metric}
	\dd s^2 = r_s^2 \left( \dd \xi^2 - \sinh^2 \! \xi \, \dd \tau ^2 \right), 
\end{align}
with $\tau \!\in\! (-\infty, +\infty), \, \xi \!\in\! [0, + \infty)$. Notice in particular that $\xi = 0$ corresponds to $r = r_s$. This is an AdS$_2$ metric. To see this explicitly, we note that these coordinates correspond to the embedding 
\begin{align}
	X_0 = r_s \cosh \xi, \qquad\qquad  X_1 = r_s  \sinh \xi \sinh \tau, \qquad\qquad  X_2 = r_s  \sinh \xi \cosh \tau,
\end{align}
which satisfy $-X_0^2 - X_1^2 +X^2_2 = -r_s^2$, and cover the region of this hyperboloid that satisfies $X_0 \geq r_s, X_2\geq 0$. This portion of AdS$_2$ is depicted in Figure~\ref{fig}.

\subsection{(Conformal) Killing vectors of near-zone Kerr geometry}

\noindent The (conformal) Killing vectors of the near-zone Kerr metric in eq.~\eqref{near zone metric Kerr} (or, equivalently, eq.~\eqref{near zone 2}) are:
\begin{subequations}
\label{app:genkerr}
\begin{align}
T &= \mathcal{R}\partial_t+\tfrac{2a}{r_\star}\partial_\varphi,\label{Tkerr}\\
J_{01} &= - \tfrac{2 \Delta}{r_\star} \cos \theta \, \partial_r - \tfrac{\partial_r \Delta}{r_\star} \sin \theta \,\partial_\theta, \label{J01kerr} \\
J_{02} &= - \cos \varphi'  \left[ \tfrac{2 \Delta}{r_\star} \sin \theta \, \partial_r + \tfrac{\partial_r \Delta}{r_\star} \left( \tfrac{\tan\varphi'}{\sin\theta} \partial_\varphi - \cos \theta \partial_\theta \right) \right], \\
	J_{03} &= - \sin \varphi' \left[ \tfrac{2 \Delta}{r_\star}  \sin \theta \, \partial_r - \tfrac{\partial_r \Delta}{r_\star} \left( \tfrac{\cot\varphi'}{\sin\theta} \partial_\varphi + \cos \theta \partial_\theta \right) \right], \\
J_{12} &= \cos \varphi' \partial_\theta - \cot \theta \sin \varphi' \, \partial_\varphi , \\
J_{13} &= \sin \varphi' \partial_\theta + \cot \theta \cos \varphi' \, \partial_\varphi, \\
J_{23} &= \partial_\varphi,\\
L_\pm &= e^{\pm t/\mathcal{R}}\left[\mathcal{R}(\partial_r\sqrt\Delta)\partial_t\mp\sqrt\Delta\partial_r+\tfrac{2a}{r_\star}(\partial_r\sqrt\Delta)\partial_\varphi\right]\label{Lkerr}\\
K_{\pm} &= e^{\pm t/{\cal R}} \tfrac{\sqrt{\Delta}}{r_\star}  \cos \theta \left( \tfrac{\rs \rp r_\star}{\Delta}\partial_t \mp \partial_r \Delta \partial_r \mp 2 \tan \theta \partial_\theta +\tfrac{ar_\star}{\Delta}\partial_\varphi \right) , \\ 
M_{\pm} &= e^{\pm t/{\cal R}} \cos \varphi'  \left[ \tfrac{\rs r_+}{\sqrt{\Delta}} \sin \theta \partial_t \mp \tfrac{\sqrt{\Delta} \partial_r \Delta \sin \theta}{r_\star } \partial_r  \pm \tfrac{2 \sqrt{\Delta}}{r_\star} \cos \theta \partial_\theta +\left(\tfrac{a\sin\theta}{\sqrt\Delta}\mp \tfrac{2 \sqrt{\Delta}}{r_\star} \tfrac{\tan \varphi'}{\sin \theta}\right) \partial_\varphi \right] , \\
N_{\pm} &= e^{\pm t/{\cal R}} \sin \varphi'  \left[ \tfrac{\rs r_+}{\sqrt{\Delta}} \sin \theta \partial_t \mp \tfrac{\sqrt{\Delta} \partial_r \Delta \sin \theta}{r_\star } \partial_r  \pm \tfrac{2 \sqrt{\Delta}}{r_\star} \cos \theta \partial_\theta +\left(\tfrac{a\sin\theta}{\sqrt\Delta}\pm \tfrac{2 \sqrt{\Delta}}{r_\star} \tfrac{\cot \varphi'}{\sin \theta}\right) \partial_\varphi \right],
\end{align}
\end{subequations}
where we have defined $\mathcal{R} = \frac{2\rs \rp}{r_\star}$, $\varphi'=\varphi-\frac{a}{\rs\rp}t$ and $r_\star\equiv r_+-r_-$. Note that these reduce to~\eqref{KVs} and~\eqref{CKVs} in the limit $a\to 0$.
The generators  \eqref{app:genkerr} satisfy the ${\rm so}(4,2)$ algebra. We can make this explicit by defining  $J_{54} = T$ along with
\begin{subequations}
\label{Ji4}
\begin{align}
	 J_{04} &=  \frac{L_+ - L_-}{2}, & J_{05} &=  \frac{L_+ + L_-}{2} , \\
   J_{14} &=  \frac{K_+ - K_-}{2}, &  J_{15} &=  \frac{K_+ + K_-}{2} , \\ 
	 J_{24} &=  \frac{M_+ - M_-}{2}, & J_{25} &=  \frac{M_+ + M_-}{2} , \\
	 J_{34} &=  \frac{N_+ - N_-}{2}, & J_{35} &=  \frac{N_+ + N_-}{2},
\end{align}
\end{subequations}
which then have the  ${\rm so}(4,2)$ commutation relations
\be
	[J_{AB}, J_{CD}] = \eta_{AD} J_{BC} + \eta_{BC} J_{AD} - \eta_{AC} J_{BD} - \eta_{BD} J_{AC} ,
\label{SO42cr}
\ee
where $\eta_{AB} = \text{diag} (-1, 1, 1, 1, 1, -1)$.
A few comments are in order. 
First, note that only $J_{01}$ and $J_{23}$ are time-independent
(when expressing all quantities in Boyer--Lindquist coordinates)
and are exact symmetries of the static sector~\cite{Hui:2021vcv}. The other generators depend  explicitly on time and are not (C)KVs  of the effective 3D Kerr metric of \cite{Hui:2021vcv}. Second, the generators $L_\pm$ differ from the ones introduced in~\cite{Charalambous:2021kcz} for non-zero values of the spin parameter $a$. Interestingly, though, they coincide (up to a rescaling of the time coordinate) in the region close to the horizon defined by $(r - r_+)/r_+ \ll 1$. This is a manifestation of the fact that all the near-zone approximations of Kerr put forward in the literature actually coincide in this limit. 
Note also that some of the generators in \eqref{app:genkerr} are manifestly well defined in the extremal limit ($a \to r_s/2$, $r_\star\rightarrow0$), while others look singular in this limit. This is not a problem because one can consistently recover all the (C)KVs of the  metric \eqref{near zone metric Kerr} at extremality by multiplying with suitable powers of $r_\star$ and taking  linear combinations  of the generators \eqref{app:genkerr}. For instance, in addition to $J_{ij}$ and $T$ (after extracting a  $1/r_\star$ factor), the other two KVs of \eqref{near zone metric Kerr} in the extremal limit are obtained by expanding $J_{04}$ and the combination  $(T -J_{05})/r_\star$ at leading order in $r_\star$.

\subsection{Ladder in spin and finite frequency}  

\noindent
A convenient near-zone approximation that describes the dynamics of particles of generic spin, $s$, in the limit $r_+ \leq r \ll1/\omega$ is \cite{Page:1976df,1974JETP...38....1S}
\begin{equation}
\label{near-zonTeuk}
x (x+1) \partial_x^2R 
+(s+1) (2 x+1)  \partial_xR
+ \left[- (\ell-s) (\ell+s+1) + \frac{q^2 +isq(2x+1)}{x(x+1)} \right]R=0,
\end{equation}
where 
$q\equiv r_sr_+(m\Omega_+-\omega)/(r_+-r_-)$ and
\begin{equation}
x\equiv \frac{r-r_+}{r_+-r_-} .
\end{equation}
It is straightforward to show that eq.~\eqref{near-zonTeuk} admits the following set of spin raising and lowering operators:
\begin{equation}
E^+_s = \left( \partial_r -\frac{  i r_sr_+ (\omega-m\Omega_+) }{\Delta } \right) R \, ,
\qquad\qquad
E^-_s  =\Delta^{-s+1} \left( \partial_r +\frac{  i r_sr_+ (\omega-m\Omega_+) }{\Delta } \right) \Delta^s R \, ,
\label{LaddersKerrPage}
\end{equation}
which generate solutions with spin $s+1$ and $s-1$ respectively, i.e., $R^{(s+1)}=E^+_sR^{(s)}$ and $R^{(s-1)}=E_s^-R^{(s)}$, where $R^{(s)}$ solves \eqref{near-zonTeuk} with spin $s$.
The operators~\eqref{LaddersKerrPage} generalize the Teukolsky--Starobinsky identities~\cite{Press:1973zz,1974JETP...38....1S,Chandrasekhar:1985kt} in the near-zone regime by connecting solutions with consecutive spin $s$ and $s\pm1$. These spin raising and lowering operators provide a simple way of extending  the results discussed above for spin-0 fields to spin-1 and spin-2 particles described by the Teukolsky equation~\cite{Hui:2021vcv}.

\end{document}